\documentclass[12pt]{article}

\usepackage[margin=1in]{geometry} 
\usepackage{lineno}
\usepackage[utf8]{inputenc}
\usepackage{hyperref}

\usepackage[font=small,labelfont=bf]{caption}
\usepackage[sort,numbers,square]{natbib}
\usepackage{graphicx, color}
\graphicspath{{fig/}}
\usepackage{lipsum}

\title{Dense suspensions as trainable rheological metafluids}
\author{Hojin Kim$^{1,2}$, Samantha M. Livermore$^{1,3}$, Stuart J. Rowan$^{2,4}$, \\ and Heinrich M. Jaeger$^{1,3}$}

\date{
	$^1$James Franck Institute, The University of Chicago, Chicago, Illinois 60637, USA\\%
  $^2$Pritzker School of Molecular Engineering, The University of Chicago, Chicago, Illinois 60637, USA\\
  $^3$Department of Physics, The University of Chicago, Chicago, Illinois 60637, USA\\
    $^4$Department of Chemistry, The University of Chicago, Chicago, Illinois 60637, USA\\ [2ex]
	\today
}
\begin{document}
	\maketitle
	
	\begin{abstract}
Memory-forming properties introduce a new paradigm to the design of adaptive materials. In dense suspensions, an adaptive response is enabled by non-Newtonian rheology; however, typical suspensions have little memory, which implies rapid cessation of any adapted behavior. Here we show how multiple adaptive responses can be achieved by designing suspensions where different stress levels trigger different memories. This is enabled by the interplay of interactions based on frictional contact and dynamic chemical bridging. These two interactions lead to novel rheology with several well-delineated shear thinning and thickening regimes, which enable stress-activated memories associated with opposite time-dependent trends. As a result, in response to different stress levels, the suspension can evolve by either softening or stiffening and is trainable, exhibiting targeted viscosity and energy dissipation with repeated low-velocity impact. Such behavior, usually associated with mechanical metamaterials, suggests that dense suspensions with multiple memories can be viewed as trainable rheological metafluids.
	\end{abstract}

\newpage

Dense suspensions exhibit non-Newtonian rheology by undergoing transitions between shear-induced microstructural states, resulting in a viscosity that can vary widely with applied stress. 
These often dramatic viscosity changes include shear thinning or thickening as well as combinations thereof \cite{Morris_2020,brown2010generality,guy2018constraint}.
In particular, discontinuous shear-thickening, whereby the viscosity increases abruptly with applied shear rate, has been explored for a range of smart, stress-adaptive materials \cite{lee2003ballistic,haris2015shear}.
To control and tailor the shear-thickening behavior of dense suspensions, studies have varied the interactions between particles by changing the particle surfaces to affect the frictional contact properties \cite{Hsu_2018,hsu2021exploring}, exploiting chemical bridging \cite{james2018interparticle,jackson2022designing,kim2023} as well as particle-solvent interactions \cite{van2021role,raghavan1997shear}, or applying external perturbations such as acoustic forces \cite{lin2016tunable,sehgal2019using} or magnetic fields \cite{brown2010generality}.

In most suspensions, the shear-induced microstructural changes do not persist when the stress is removed and, except for brief transients, there is no memory of the shearing history. 
Sustained memory requires interactions that can persist, effectively generating feedback between the microstructural outcome and the applied shear that produces this outcome (Fig.~\ref{fig:fig1}A). 
Such memory can give rise to more complex, time-dependent rheologies like thixotropy, where the suspension viscosity decreases over time, or rheopexy (anti-thixotropy), which exhibits the opposite trend \cite{mewis1979thixotropy}.
However, suspensions typically exhibit only a single regime of stress-adaptive shear thickening or a single type of memory. 
A new and conceptually different situation arises when the particle interactions are designed to enable several, independently stress-adaptive behaviors and when each of these behaviors can furthermore retain a different memory of the shearing history (as indicated by the switches in Fig.~\ref{fig:fig1}A). 
Suspensions with multiple memories of this kind exhibit rheological characteristics that could be considered metamaterial properties, albeit for a fluid; hence, suitably engineered suspensions can be thought of as \emph{rheological metafluids}.
We define rheological metafluids as distinct from other classes of metafluid, such as those designed to have tunable mass density \cite{bi2023acoustic} or optical properties \cite{djellouli2024shell}, and from metamaterials whose viscosity can be changed through external excitation \cite{sehgal2024programmable}.

Here we demonstrate such metafluid behavior with suspensions designed to have two shear-activated memories that control opposite rheological behaviors in the same system. 
One type of memory is activated at low stress and associated with the growth of larger particle aggregates, which increases the viscosity and thereby controls rheopectic behavior. 
The other memory is activated at higher stress levels and is associated with the destruction of particle aggregates, which decreases the viscosity and controls thixotropic behavior. 
The result is a novel combination of tailorable rheological properties, including dual shear thickening regimes, rheopexy, and thixotropy, all within a single fluid but across different stress ranges.
These memories are embedded in the shear-induced microstructure of the suspension and this makes it possible to train-in associated properties, thereby achieving different outcomes through suitable protocols of applied shear. 
We show this multifunctional capability by training-in different impact resistance and energy absorption properties through repeated mechanical impact at different velocities.

\section*{Design and shear rheology of the metafluid}	

\begin{figure}
\centering
	\includegraphics[width = 1\columnwidth]{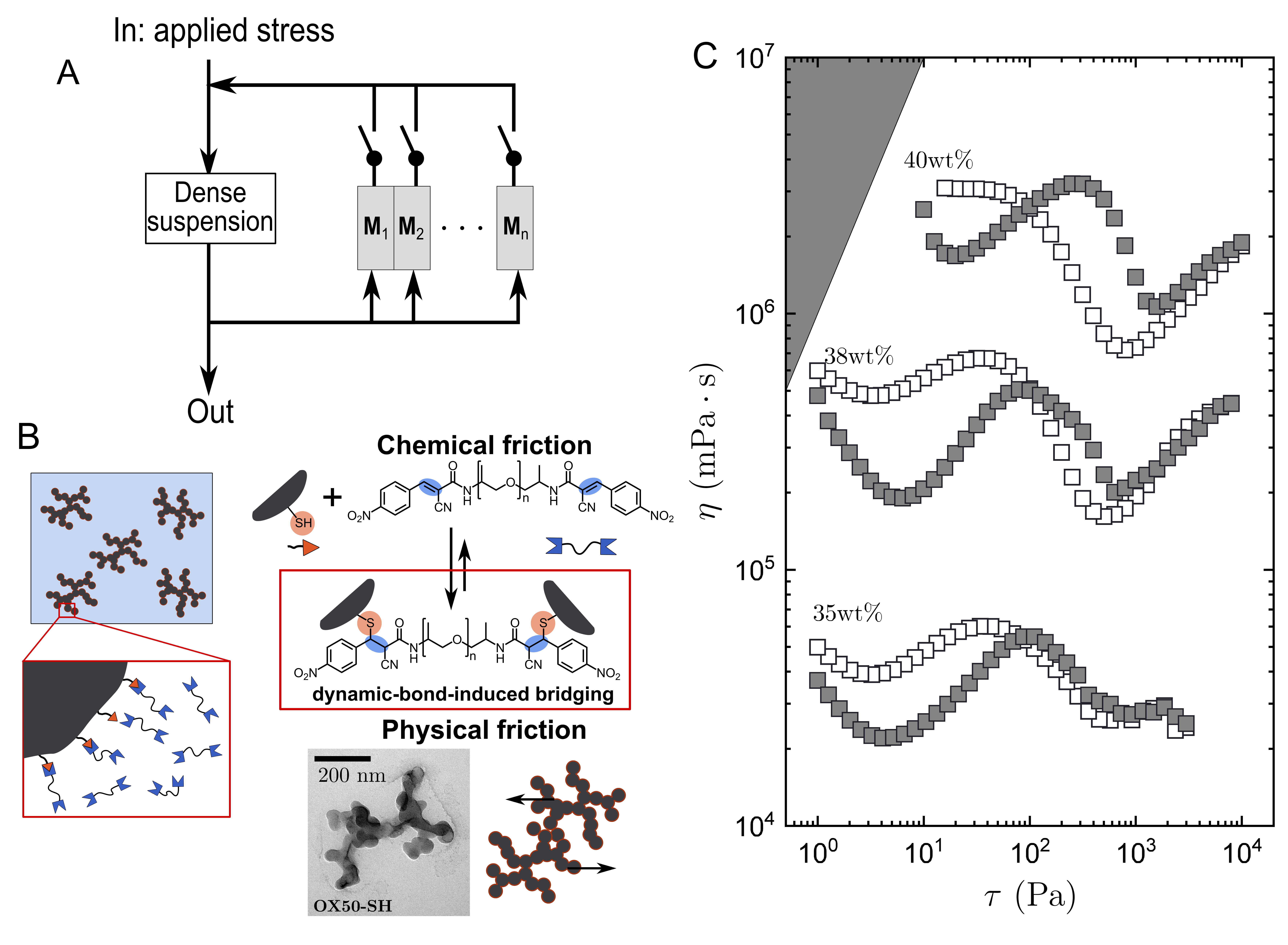}
\caption{\label{fig:fig1} \textbf{Design and rheology of trainable suspensions with multiple memories.} 
(A) Trainable suspensions can achieve varying viscosity states through training protocols. These protocols enable memory formation (\textbf{M}), allowing for an extended lifetime of shear-induced microstructural states. Introducing multiple memory-forming units (\textbf{M}\textsubscript{1}, \textbf{M}\textsubscript{2}, \ldots, and \textbf{M}\textsubscript{n}) enables trainable material properties, thereby tailoring viscosity states. The trained states are persistent until they relax back. (B) Schematic illustration of the metafluid design. Fumed silica particles are surface-functionalized with thiol groups (OX50-SH, red triangles) and suspended in a NO\textsubscript{2}-BCAm-endcapped poly(propylene glycol) Michael-acceptor (concave pentagons). Two distinct interparticle interactions are induced under shear: 1) chemical friction via the thia-Michael reaction between surface thiols and NO\textsubscript{2}-BCAm motifs, and 2) particle-particle frictional force. The inset shows a transmission electron microscopy image of thiol-functionalized fumed silica particles. (C) Rheology of the OX50-SH in NO\textsubscript{2}-BCAm-endcapped poly(propylene glycol) Michael-acceptor suspensions
at weight fractions $\phi_ \mathrm{w}=35$, 38, and 40 \% (bottom to top). Viscosity $\eta$ is measured by the forward (closed symbol) and backward (open symbol) ramp of applied shear stress $\tau$. The shaded area is the inaccessible window that reaches minimum shear rates measurable by the rheometer.}
\end{figure}

For the design of the metafluid suspension, we exploit the interplay of distinct physical and chemical particle-particle interactions: strong friction at direct contact and molecular bridging via dynamic covalent bonds (Fig.~\ref{fig:fig1}B).
Fumed silica particles (AEROSIL OX 50, Evonik) are used because of their rough non-spherical shape, resulting from random aggregation of $\sim$40~nm diameter primary silica spheres, which allows for highly frictional interactions and facilitates the formation of large particle clusters under shear.
The particles are surface-functionalized with thiol groups (OX50-SH) and suspended in ditopic \textit{p}-nitrobenzalcyanoacetamide (NO\textsubscript{2}-BCAm)-endcapped poly(propylene glycol) macromonomers. 
A thia-Michael reaction between the surface thiol and the NO\textsubscript{2}-BCAm motif can then generate room temperature, catalyst-free dynamic bonds between particles when they come into close proximity \cite{herbert2020dynamic,fitzsimons2021effect,jackson2022designing,kim2023}.

The rheology of OX50-SH particles suspended in NO\textsubscript{2}-BCAm-endcapped poly(propylene glycol) was measured using a parallel-plate geometry (Anton-Paar MCR-301 and MCR-702 rheometers).
Fig.~\ref{fig:fig1}C shows the viscosity as a function of shear stress, measured by ramping the shear stress $\tau$ from low to high and then back down again (10 s at each stress), for three different particle weight fractions $\phi_\mathrm{w}$. 
During the ramp-up, well-delineated shear thickening and thinning regimes appear in succession.
At the very lowest stresses, fumed silica particles form weakly coupled clusters or flocs \cite{raghavan2000rheology}.
When these flocs break under increased applied stress, the suspension shear thins and the viscosity drops.
However, as particles become sheared into close proximity and start to form aggregates through dynamic bridging bonds, the suspension transitions to shear thickening \cite{jackson2022designing, kim2023}.
Note that bridging interactions can start at significantly lower stress than direct contact because the macromonomer can span finite interparticle distances \cite{kim2024shear}.

Beyond some maximum sustainable stress level, the molecular bridges are sheared apart and the viscosity drops again.
This shear thinning gives way to a second thickening regime once the applied stress becomes sufficiently large to produce direct, frictional contact between particles \cite{guy2015towards,kim2024shear}.
Finally, as those frictionally stabilized aggregates are broken up at the largest stresses, the viscosity starts to turn around yet again into a secondary shear thinning behavior (visible most clearly at lower particle concentration). 

Comparing the curves for the forward and backward stress ramps in Fig.~\ref{fig:fig1}C, the traces differ strikingly and cross each other around $\tau \approx$ 100 Pa.
Starting from the largest stresses, as $\tau$ is ramped down below the level where aggregates can again form via bridging bonds, i.e., below $\approx$ 1000 Pa, gradual build-up of such bridged aggregates increases the viscosity. 
Because of the mildly thixotropic behavior associated with direct contact \cite{dullaert2005thixotropy,larson2019review}, the viscosity increase on ramping down starts from a value below that at the onset of the second shear thickening regime.
This produces hysteresis and eventually causes the curves to cross over one another when the backward-ramp viscosity exceeds the values of the forward-ramp in the first shear thickening regime.

The aforementioned memory-forming capabilities emerge from the time-dependent rheology on either side of the crossing of the forward and backward viscosity curves.
In the shear stress range 100 Pa $<\tau<$ 1000 Pa, the viscosity of the backward ramp is lower than that of the forward ramp and the behavior is thixotropic.
In the range 10 Pa $<\tau<$ 100 Pa this trend is reversed, yielding a higher viscosity for the backward ramp and the behavior is rheopectic. 
To elucidate the different time-dependencies, the viscosity is traced over time at constant shear rate $\dot{\gamma}$, encompassing both the low-$\dot{\gamma}$ rheopectic and the high-$\dot{\gamma}$ thixotropic regime (see Supplementary Information, Fig. S1, for the suspension with $\phi_\mathrm{w}=40$ \%). At low $\dot{\gamma}\leq 0.3$ s\textsuperscript{-1}, where the suspension exhibits rheopectic rheology, $\eta$ increases over time, similar to silica spheres in ditopic macromonomers forming dynamic covalent bonds \cite{kim2023,jackson2022designing}.
By contrast, the trend of viscosity at larger $\dot{\gamma}=0.5$, 1, and 3 s\textsuperscript{-1} shows a distinctive decrease in $\eta$ over time, i.e., thixotropic rheology.

When the concentration of particles is decreased from $\phi_\mathrm{w}=40$~\% to 35~\%, the direct-contact-induced shear thickening at large $\tau$ and the associated thixotropy diminish (Fig.~\ref{fig:fig1}C), which we associate with the break-up of sufficiently large aggregates that can be stabilized by physical frictional interactions.
When the thia-Michael bridging interaction is not present in unfunctionalized fumed silica particles, only a single shear-thickening regime at large applied stress is observed together with mild thixotropy (see Supplementary Information, Fig. S2). 
Taken together, this demonstrates that the rheopectic behavior on the low-stress side of the crossing is controlled by the dynamic-covalent-bond-induced bridging interaction, while the thixotropic behavior on the high-stress side is controlled by the interplay between this bridging and the thixotropy inherent to contact interactions in fumed silica suspensions.

\begin{figure}[!t]
\centering
	\includegraphics[width = 0.8\columnwidth]{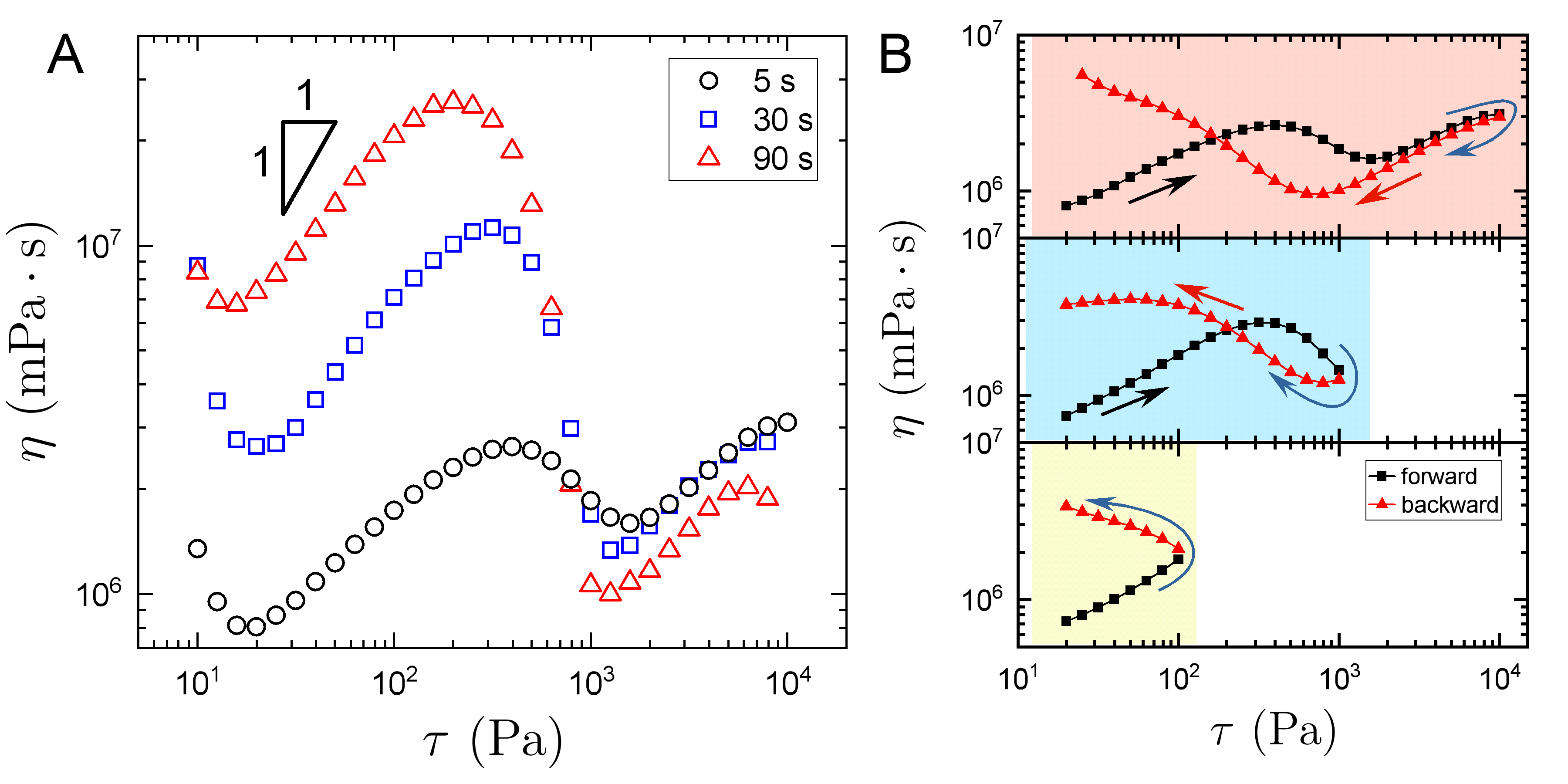}
\caption{\label{fig:hysteresis} \textbf{Hysteresis and time-dependence of shear rheology.}
(A) Viscosity of the $\phi_ \mathrm{w}=40$ \% suspension in Fig.~\ref{fig:fig1}C for different stress-controlled ramp rates, where 5 s (circles), 30 s (squares) or 90 s (triangles) were allotted to measuring each data point. 
Only forward ramps are shown.
(B) Hysteresis loops for different stress ranges. 
The colored area is the range of $\tau$ measured. Arrows indicate the direction of the stress sweep.
}
\end{figure}

The low-stress rheopectic behavior originates from the increasing number of bridging bonds over time \cite{jackson2022designing,kim2023}. 
This is shown by varying the ramp rate, i.e., varying the time allotted for each $\tau$ measurement. Figure~\ref{fig:hysteresis}A shows the change in viscosity when this time is increased from 5~s to 90~s.
The result is a large upward shift in viscosity in the low-$\tau$ regime around 100 Pa, while the shear thickening behavior changes very little and the slope $\frac{\mathrm{d} \log{\eta}}{\mathrm{d}\log{\tau}}\sim 1$, where a slope of 1 implies discontinuous shear thickening. 
By contrast, the viscosity in the high-$\tau$ regime above 1000 Pa remains unaffected (within the range of repeatability from run to run), confirming the independent origins for the dual shear thickening. 
As the ramp rate is decreased, these trends in combination create an increasingly drastic thinning behavior between the two thickening regimes due to the increased number of bridging bonds that are broken in this $\tau$ range. 
Importantly, the critical onset stress for shear thickening remains essentially the same for both regimes, indicating that it is the critical stress that leads to the thickening behavior.

By changing the maximum $\tau$ in the stress-controlled ramps, the rheopectic and thixotropic rheologies can be activated separately. We demonstrate this in Fig.~\ref{fig:hysteresis}B by ramping forward and then backward over three different $\tau$ ranges. For $10\ \mathrm{Pa}<\tau< 10^2\ \mathrm{Pa}$ (bottom panel), only rheopexy appears but thixotropy is absent. 
The thixotropic behavior can be switched on by extending the range of applied stress to $\tau=10^3\ \mathrm{Pa}$ (middle).
Extending this to the full range $10\ \mathrm{Pa}<\tau<10^4\ \mathrm{Pa}$ (top) intensifies the thixotropic viscosity drop on ramping backward, but, as expected from physical friction-induced thickening at large stress, does not show significant hysteresis (only the very slight thixotropy associated with unfunctionalized fumed silica). Interestingly, the crossing stress of the forward and backward viscosity ramps is independent of the measured stress range. This again demonstrates that the rheopectic and thixotropic behaviors are independent and, therefore, function as distinct memory units.
 
\section*{Training different memories}

\begin{figure}[!ht]
\centering
	\includegraphics[width = 0.8\columnwidth]{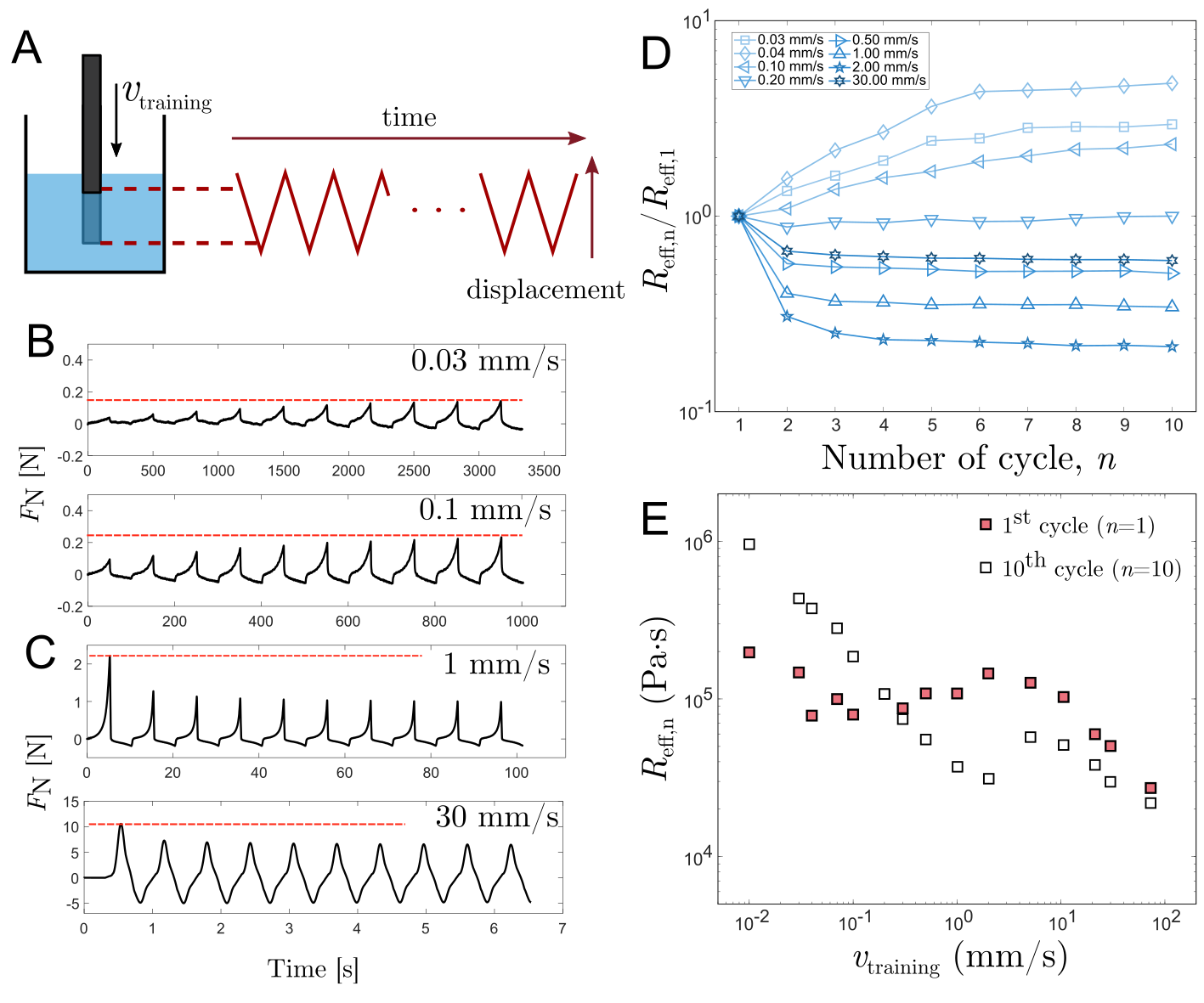}
\caption{\label{fig:impact} \textbf{Training via repeated vertical impact.} (A) Setup for the impact test. For the training process, the impacting rod travels $L=5$~mm up and down at a fixed velocity $v_\mathrm{train}$ and the normal force $F_\mathrm{N}$ on the impactor is tracked as a function of time. (B) For low $v_\mathrm{train}$, stiffening of the suspension is observed. (C) The suspension softens when repeated impact occurs with larger $v_\mathrm{train}$. 
(D) Effective impact resistance $R_\mathrm{eff,n}$ during training as a function of training cycle $n$, normalized by $R_\mathrm{eff,1}$ for the first cycle. 
(E) $R_\mathrm{eff,n}$ for the first cycle ($n=1$, closed symbols) and the final cycle ($n=10$, open symbols) as a function of $v_\mathrm{train}$. 
Red horizontal lines in (B) and (C) refer to the maximum $F_\mathrm{N}$ during the 10 training cycles.}
\end{figure}

The ability of shear-induced microstructural states to retain a memory of the applied stress makes it possible to evolve the suspension’s properties in a desired direction, i.e., to train the metafluid. 
In the following, we discuss how applying a repeated stress loading as a training stimulus allows us to tailor the mechanical properties of the suspension such as its effective impact resistance ($R_\mathrm{eff}$). 
We demonstrate this by applying low-velocity impacts to the free surface of the suspension with particle concentration $\phi_\mathrm{w}=40$~\%. 
An impactor (Zwick-Roell Z1.0 universal tester, equipped with a 1~kN load cell) pushes a cylindrical rod (6.4~mm diameter) vertically into the suspension to a fixed depth $L=5$~mm, measured from the free surface at rest (Fig.~\ref{fig:impact}A).
This is done for 10 cycles under speed-controlled conditions using a wide range of different training velocities $v_\mathrm{train}$, while the impactor records the normal force on the rod. 
Depending on the impact speed $v_\mathrm{train}$ during training, we find opposite behaviors.
For sufficiently low speeds, the peak magnitude $F_\mathrm{N}$ of the normal force at the $n$-th impact increases over successive cycles.
This is shown in Fig.~\ref{fig:impact}B for two example traces and implies a gradual stiffening of the suspension.
However, training at larger speed $v_\mathrm{train}$ leads to a softening, as seen in Fig.~\ref{fig:impact}C by the decrease in the peaks of $F_\mathrm{N}$. 

The effect of such training can be quantified by estimating the effective impact resistance $R_\mathrm{eff,n}$ resulting from each successive impact. 
We take $R_\mathrm{eff,n}=\tau_\mathrm{eff}/\dot{\gamma}_\mathrm{eff}$ with an effective stress given by $\tau_\mathrm{eff}=F_\mathrm{N}/A$, where $A$ is the surface area of the impacting rod (6.4~mm in diameter), and an effective shear rate given by $\dot{\gamma}_\mathrm{eff}=v_\mathrm{train}/L$. 
For the calculation of $R_\mathrm{eff,n}$ we use $F_\mathrm{N}$ at a displacement of 4~mm, i.e., 1~mm before the full depth of the impact stroke is reached, in order to exclude boundary effects. 
In Fig.~\ref{fig:impact}D, $R_\mathrm{eff,n}$ normalized by the initial, untrained resistance $R_\mathrm{eff,1}$ is plotted for different training speeds.
For $v_\mathrm{train}<0.2$~mm/s, the data show a gradual viscosity increase, which saturates for $n>6$. 
For example, low-velocity training at $v_\mathrm{train}=0.04$~mm/s increases the fluid resistance up to 4 times compared to the untrained case $R_\mathrm{eff,1}$. 
In contrast, training at $v_\mathrm{train}>0.5$~mm/s leads to a decrease in resistance. 
For instance, training at $v_\mathrm{train}=2$~mm/s produces a resistance almost 5 times lower than $R_\mathrm{eff,1}$. 

We can tie this training-induced behavior of the metafluid to its shear rheology by taking the crossing of the forward and backward viscosity ramps in Fig.~\ref{fig:fig1}C to estimate the impact speed that separates stiffening from softening.
For the $\phi_\mathrm{w}=40$~\% suspension, this crossing occurs at a stress of about 100 Pa and a viscosity of $3\times10^6 \ \mathrm{mPa\cdot s}$, corresponding to a shear rate $\tau/\eta = (1/3) \times 10^{-1} \ \mathrm{s^{-1}}$.
Equating this with the effective rate $\dot{\gamma}_\mathrm{eff}=v_\mathrm{train}/L$ during training with $L=5$~mm, we obtain that the crossing seen in the shear rheology corresponds to an impact speed of approximately 0.2~mm/s.
Since the crossing point in the shear rheology defines conditions under which the time-dependent behavior vanishes, we can expect minimal evolution of the effective impact resistance if the training occurs at the corresponding impact speed.
In Fig.~\ref{fig:impact}D we indeed find minimal changes in $R_\mathrm{eff,n}$ with $n$ when $v_\mathrm{train} \approx 0.2$~mm/s. 
This is highlighted in Fig.~\ref{fig:impact}E, which compares the initial resistance $R_\mathrm{eff,1}$ (open symbols) with the trained value $R_\mathrm{eff,10}$ after 9 additional impact cycles (closed symbols).
The correspondence between the shear rheology (Fig.~\ref{fig:fig1}C) and the resistance evolution during impact training at different speeds (Fig.~\ref{fig:impact}E) demonstrates that the trainability originates from the time-dependent rheology of the metafluid and the associated memories.

Note that the degree of trainability, given by the difference between the open and closed symbols in Fig.~\ref{fig:impact}E, is reduced significantly when the impact speed exceeds values around 10~mm/s, which corresponds to the onset of the second shear thickening regime for the $\phi_\mathrm{w}=40$~\% suspension.
We do not expect much trainability in that regime because the thixotropy in the stress range dominated by contact friction is minimal. 
This is verified by impacting suspensions made from the same fumed silica particles but suspended in a liquid that does not form dynamic bridging bonds and where therefore shear thickening is solely due to contact friction (see Fig. S3 for a fumed silica suspension suspended in poly(ethylene glycol) 200~g/mol).
In this case, $R_\mathrm{eff,n}$ does not evolve over successive training cycles and instead keeps a fixed value.

\begin{figure}[!t]
\centering
	\includegraphics[width = 1\columnwidth]{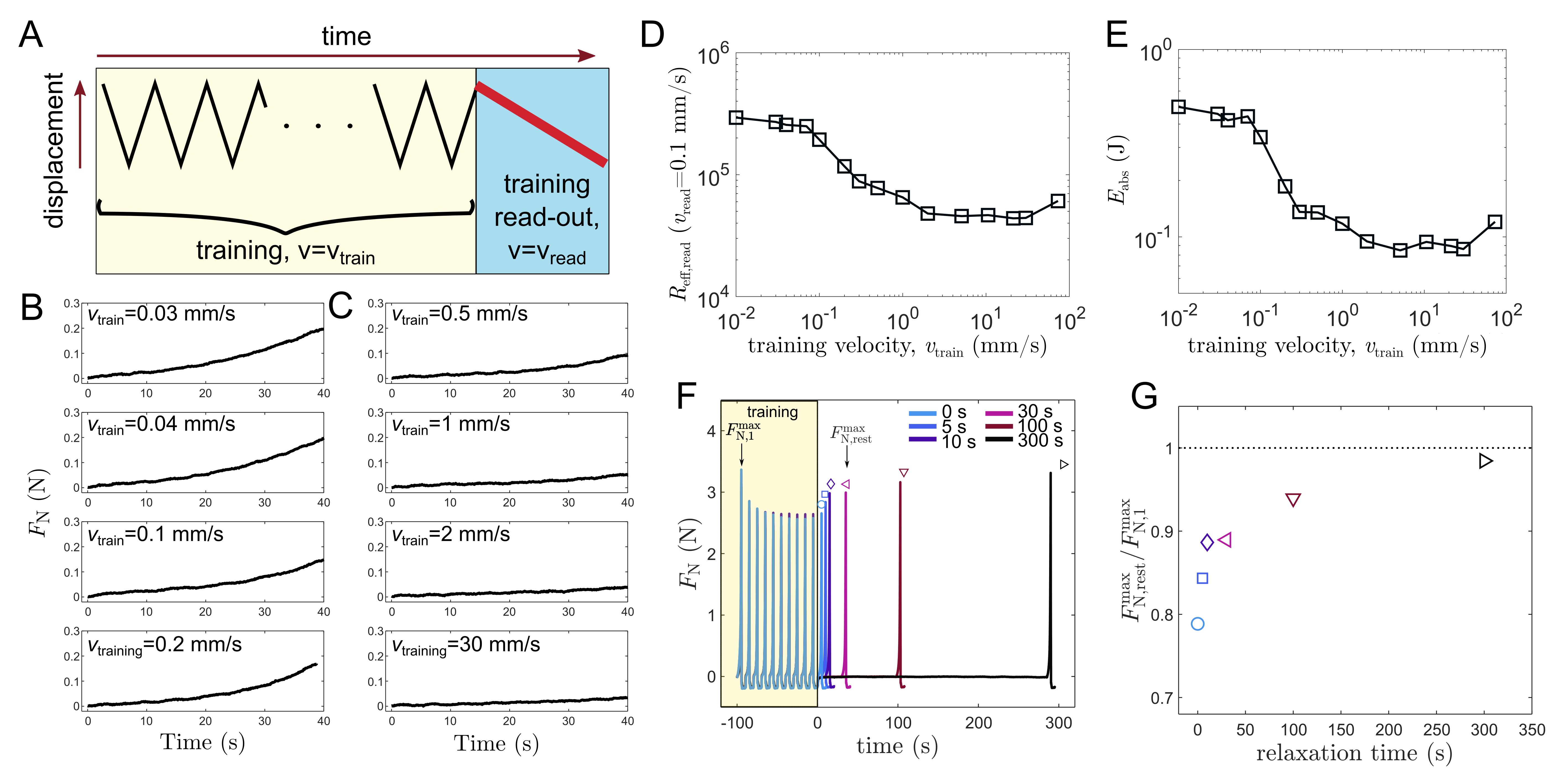}
\caption{\label{fig:readout} \textbf{Read-out of memories in the trained metafluid.} (A) Illustration of the read-out protocol (blue background) after 10 cycles of training. After training at given $v_\mathrm{train}$ (yellow background), the impactor rod compresses the trained fluid at fixed $v_\mathrm{read}=0.1$~mm/s while the normal force $F_\mathrm{N}$ is measured. (B) Read-out force for low impact speeds $v_\mathrm{train}=0.03$, 0.04, 0.1, and 0.2~mm/s (top to bottom).
(C) Read-out force for larger impact speed $v_\mathrm{train}$= 0.5, 1, 2, and 30~mm/s (top to bottom). (D) Effective impact resistance during read-out $R_\mathrm{eff,read}$ at the impact depth $L=4$~mm and (E) energy absorbed $E_\mathrm{abs}$ during read-out with constant read-out velocity $v_\mathrm{read}=0.1$~mm/s after being trained at varying training velocities $v_\mathrm{train}$.
(F) Lifetime of memories trained at $v_\mathrm{train}=1$~mm/s. The mechanical response of the trained suspension is measured at the same compression speed of 1~mm/s after a relaxation time of 0, 5, 10, 30, 100, and 300 s. (G) Peak normal force during read-out $F^\mathrm{max}_\mathrm{N,read}/F^\mathrm{max}_\mathrm{N,1}$ normalized by the peak force of the untrained suspension $F^\mathrm{max}_\mathrm{N,1}$ as a function of relaxation time.
}
\end{figure}

Many applications of dense suspensions focus on the energy dissipation properties. 
In this regard, trainability makes it possible to tailor the mechanical response through different training protocols. 
We examine this through a two-step process, whereby we first train the suspension through 10 impact cycles to imprint a memory.
We then immediately switch to probing the trained-in impact response with a separate `read-out' compression at fixed velocity $v_\mathrm{read}=0.1$~mm/s (Fig.~\ref{fig:readout}A). 
 Figure~\ref{fig:readout}B shows representative traces of $F_\mathrm{N}$ as a function of time during read-out at $v_\mathrm{read}=0.1$~mm/s, after being trained at low $v_\mathrm{train}=0.03$, 0.04, 0.1, and 0.2~mm/s (top to bottom). A decrease in $F_\mathrm{N}$ during the read-out compression is seen as $v_\mathrm{train}$ increases. The most drastic decrease is observed for larger training velocities (Fig.~\ref{fig:readout}C) using $v_\mathrm{train}=0.5$, 1, 2, and 30~mm/s (top to bottom). Fig.~\ref{fig:readout}D illustrates the range of $R_\mathrm{eff,read}$ (measured at impact depth $L=4$~mm) changes resulting from different training conditions, reflecting the range in achievable mechanical response. The effective impact resistance measured at the read-out step $R_\mathrm{eff,read}$ shows that as a result of training, the final trained resistance can be controlled significantly, ranging from $R_\mathrm{eff}\sim 5 \times 10^{4}$ to $3\times 10^{5}\ \mathrm{Pa\cdot s}$.

Changes in mechanical stiffness directly influence the magnitude of energy dissipation.
The absorbed energy $E_\mathrm{abs}$ during the read-out impact is estimated by integrating the work done: $E_\mathrm{abs}=\int F_\mathrm{N}(L) \,dL$ during the first 80\% of impact to avoid the boundary effect near the wall. 
The total amount of energy absorbed at a low-speed read-out $v_\mathrm{read}=0.1$~mm/s is proportional to the material $R_\mathrm{eff}$ after the training (Fig.~\ref{fig:readout}E). A stiffened material trained at a slow velocity has a 5$\times$ greater impact-mitigating property than the high-velocity-trained fluid. 
The mechanical response of trained suspensions can be influenced by the impact speed depending on the lifetime of memory and its resilience. 
To elucidate how the trained materials respond to different impact events, we explore the amount of energy absorption $E_\mathrm{abs}$ at a higher impact speed $v_\mathrm{read}=5$~mm/s (50 times faster impact) following the same training procedure (see supplemental materials, Fig. S4). 
The overall dissipated energy is increased by one order of magnitude owing to a higher speed impact at read-out. 
Interestingly, the memory in the fluid is still persistent under a fast impact. 
For a fluid trained at low-velocity ($v_\mathrm{train}<0.1$~mm/s), there is no training effect at 5~mm/s read-out velocity. 
At higher $v_\mathrm{train}>0.1$~mm/s, a systematic softening appears, reducing $E_\mathrm{abs}$, followed by a plateau at high $v_\mathrm{train}>10$~mm/s. 
As a result, the amount of mitigation can be controlled by a factor of 2.5 (from 4 to 10 J).

The versatility of training-in different properties is critically influenced by the lifetime of the material's memory. 
To discern the lifetime of trained memory, we measure the recovery of $F_\mathrm{N}$ of a trained suspension at varied relaxation times between the training and read-out steps. 
The metafluid ($\phi_\mathrm{w}=40$ \%) is trained at impact speed of 1~mm/s for 10 cycles and then left to rest for different relaxation times from 0 to 300~s. 
After relaxation, the normal force is measured during a read-out step (Fig.~\ref{fig:readout}F). 
The peak force $F^\mathrm{max}_\mathrm{N,read}$ has its extreme value right after training and then relaxes over time.
For the example in Fig.~\ref{fig:readout}F, where the training was done at large velocity, this peak force is smallest immediately after training and for longer relaxation times increases back to the value of the untrained state.
This relaxation of the trained fluid is quantified by plotting $F^\mathrm{max}_\mathrm{N,read}$ normalized by the peak normal force of the untrained suspension $F^\mathrm{max}_\mathrm{N,1}$ (Fig.~\ref{fig:readout}G). 
The ratio $F^\mathrm{max}_\mathrm{N,read}/F^\mathrm{max}_\mathrm{N,1}$ shows that the trained fluid recovers approximately 95\% of its untrained state in 100~s. This relaxation can occur without any physical agitation, facilitating a reprogrammable impact-mitigating capability upon further training.

\section*{Conclusions}
In this study, we introduce a new class of suspension-based complex fluids with multiple stress-activated memories, which behave as metafluids. 
The rheological metafluid is designed to exhibit two distinct memories in dense suspensions originating from the presence of both chemical and physical particle-particle interactions. We show that this combination makes it possible to engineer stress-adaptive rheological behavior that includes dual shear thickening regimes, rheopexy, and thixotropy. Furthermore, reinforcing the memory of shear-induced microstructure through training, i.e., repeated load application, enables this metafluid to exhibit distinct stress responses, such as stiffening or softening due to impact, depending on the training protocol.

Such trainable adaptivity differs from the behavior of materials that harden under mechanical force through mechanochemical polymerization and crosslinking mechanisms \cite{wang2021bio,wang2019mechanically,matsuda2019mechanoresponsive}. In those materials, the hardening is driven mainly by chemical reactions that are irreversible, which effectively corresponds to a single memory of infinite duration. This preserves the materials' properties but prevents them from being retrainable and often also reprocessable. By contrast, the suspension-based metafluids introduced here have multiple memories of finite duration. As with biological systems such as muscle, finite memory requires maintenance, which typically will occur through continual exposure to stresses provided by the environment. At the same time, a finite memory enables re-training for different properties \cite{Jaeger2024training}.
This feature also highlights how trainable metafluids can change their mechanical behavior in response to changes in the external environment. The lifetime of the trained-in memory and the range of trainable mechanical properties of these suspensions could be tuned further by tailoring the bond strength of the chemical interactions \cite{zhong2013studies,jackson2022designing,crolais2023enhancing}, the molecular weight of the telechelic molecules \cite{kim2023}, and the viscosity of the suspending medium, thereby expanding the phase space for customization of memory-forming metafluids.


	\newpage
 \section*{Data availability}
 The data that support the findings of this study are available within the article and Supplementary Information.

\section*{Author contributions}
H.K. and H.M.J. conceived the idea. H.K. synthesized and prepared the samples. H.K. and S.M.L performed the impact tests. All authors contributed to the analysis. H.K. and H.M.J. wrote the manuscript. All authors reviewed the manuscript.

\section*{Competing interests}
The authors declare no competing interests.

\section*{Acknowledgements}
 The authors acknowledge support from the University of Chicago Materials Research Science and Engineering Center (MRSEC), which is supported by the National Science Foundation under award DMR-2011854. This work was partially supported by National Science Foundation under award DMR-2104694.

\section*{Methods}
\subsection*{Synthesis of \textit{p}-nitrobenzalcyanoacetamide-endcapped poly(propylene glycol)}

NO\textsubscript{2}-BCAm-endcapped poly(propylene glycol) Michael-acceptor was synthesized following the previously reported method\cite{jackson2022designing}. First, 40 g (10 mmol) of Jeffamine D-4000 (Huntsman, average molecular weight of 4000 g/mol) and 2.2 stoichiometric ratio cyanoacetic acid (Alfa Aesar, 22 mmol) were added to a 250 mL round-bottom flask with a solution of 40 mL dimethylformamide (DMF) and 120 mL toluene mixture. 0.4 g of \textit{p}-toluenesulfonic acid (Sigma-Aldrich, $\geq 98.5\%$) was added to the flask. While stirring, nitrogen gas was purged for 30 min. The reaction proceeded at 120$^\circ$C for 8~hr, and the produced water was removed from the toluene and water azeotrope mixture through a Dean-Stark trap.
After the reaction, the toluene solvent was removed from the solution using a rotary evaporator, and 50 mL of chloroform was added to dilute the product. The cyanoacetamide-terminated poly(propylene glycol) (\textbf{1}) solution was washed three times with sodium bicarbonate aqueous solution and water. The washed product \textbf{1} was dried with magnesium sulfate and filtered. Finally, the remaining chloroform was removed using a rotary evaporator.

Product \textbf{1} was further reacted with 4-nitrobenzaldehyde (Sigma-Aldrich, $98 \%$) to synthesize ditopic Michael-acceptor. 10 mmol \textbf{1} and 24 mmol 4-nitrobenzaldehyde were added to a 250 mL flask with a mixture of 120 mL toluene and 40 mL DMF. 0.4 g of piperdinium acetate was added to the flask. The reaction proceeded at 120$^\circ$C for 8~hr. The same washing protocols were used to remove the unreacted reagents and piperdinium acetate. 

\subsection*{Synthesis of thiol-functionalized fumed silica particles}
Fumed silica particles (AEROSIL OX 50) were purchased from Evonik. The particle surface area is 50 m\textsuperscript{2}/g. The particles were functionalized by using the procedure reported previously\cite{crucho2017functional}. First, 10 g of silica particles were added to 400 mL of toluene. Before the reaction, the suspension was homogenized by sonicating and stirring for 1 hr. Then 10 molecules of 3-mercaptopropyl trimethoxysilane (MPTMS, Sigma-Aldrich, $95\%$) per nm\textsuperscript{2} surface area of the fumed silica particles were added to the solution, and the suspension was left at reflux for 24 hr. During the reaction, the suspension was stirred with a magnetic stirrer. After the reaction was completed, toluene was removed using a rotary evaporator. The dried powder was washed with ethanol three times by repeated cycles of sonication and centrifugation. The final particles were completely dried under vacuum for at least 24 hr. 

The density of the surface thiol functional groups was characterized with nuclear magnetic resonance spectrometry, following the method presented by Crucho \textit{et al.}\cite{crucho2017functional}. Briefly, 5~mg of particles were dissolved in 0.5~mL D2O solution (0.5~M NaOD) with 3.3~mg 1,3,5-trioxane used as an internal standard. The solution was heated to 90\textsuperscript{$\circ$}C and stirred for approximately 2 hours. The surface thiol density was determined by \textsuperscript{1}H NMR (Bruker Avance nanobay III HD 400 MHz nuclear magnetic
resonance spectrometer) $\mathrm{(ONa)_3SiC\mathbf{H_2}CH_2CH_2SH}$ peak at $\delta=0.62$~ppm, yielding 1.5 thiol-per-nm\textsuperscript{2} surface area.

\subsection*{Sample preparation and shear rheology measurement}
Suspensions were prepared by mixing dried thiol-functionalized fumed silica particles and \textit{p}-nitrobenzalcyanoacetamide-endcapped poly(propylene glycol). The suspension was mixed homogeneously and sonicated for approximately 1 hour. 

The shear rheology of the suspensions was measured using a MCR301 rheometer (Anton Paar) with a 25~mm diameter parallel plate equipped with a Peltier plate. The viscosity was measured with stress-controlled sweep measurements. The equilibrium time at each shear stress was varied to confirm the hysteretic rheological trends. 

\subsection*{Impact test}
Impact tests were conducted using a Zwick-Roell Z1.0 mechanical tester, equipped with a 1~kN load cell.
The sample was trained and impacted using a cylindrical impacting rod with a diameter 6.35~mm. 
The sample was loaded into the cylindrical beaker with a diameter of 3.5~cm and a height of 6~cm.
Experiments begin with the bottom face of the impactor rod just below the surface of the suspension to avoid introducing air bubbles during compression.
While the impactor travels through the suspension, the mechanical tester continuously records the normal force, position, and time.
For cyclic training experiments, the rod pushes into the suspension at a fixed rate up until a set depth before lifting back up to its starting height; this lowering-raising action is repeated $n$ times during training.

\subsection*{Relaxation experiments with impact test}
Relaxation experiments were performed using a Zwick-Roell Z1.0 mechanical tester (1 kN load cell). The impactor rod was submerged into the 40 wt \% suspension. 10 cycles of axial oscillation were applied to the suspension with the constant velocity $v_\mathrm{training}=1$ mm/s. The trained suspension was left for a certain relaxation time and then impacted with the same 1 mm/s velocity. To explore different relaxation times, the trained suspension was left to equilibrate and then re-trained for a different relaxation time (0 to 300 s).

\newpage

\graphicspath{{fig/}}

\renewcommand{\thetable}{S\arabic{table}}
\renewcommand{\thefigure}{S\arabic{figure}}
\renewcommand{\theequation}{S\arabic{equation}}
\renewcommand{\thesection}{S\arabic{section}.}
\renewcommand{\bibnumfmt}{S}
\setcounter{figure}{0}
\setcounter{section}{0}


\begin{center}
\Large{\textbf{Supplementary Information for\\Dense suspensions as trainable rheological metafluids}} \\
\vspace{0.2cm}

\large{Hojin Kim$^{1,2}$, Samantha M. Livermore$^{1,3}$,  Stuart J. Rowan$^{2,4}$, \\ and Heinrich M. Jaeger$^{1,3}$}
\vspace{0.2cm}

\noindent
\normalsize{$^1$James Franck Institute, The University of Chicago, Chicago, Illinois 60637, USA\\%
  $^2$Pritzker School of Molecular Engineering, The University of Chicago, Chicago, Illinois 60637, USA\\
  $^3$Department of Physics, The University of Chicago, Chicago, Illinois 60637, USA\\
    $^4$Department of Chemistry, The University of Chicago, Chicago, Illinois 60637, USA\\ [2ex]}
\end{center}
 \vspace{0.3cm}

\newpage
\begin{figure}
\centering
	\includegraphics[width = 0.5\columnwidth]{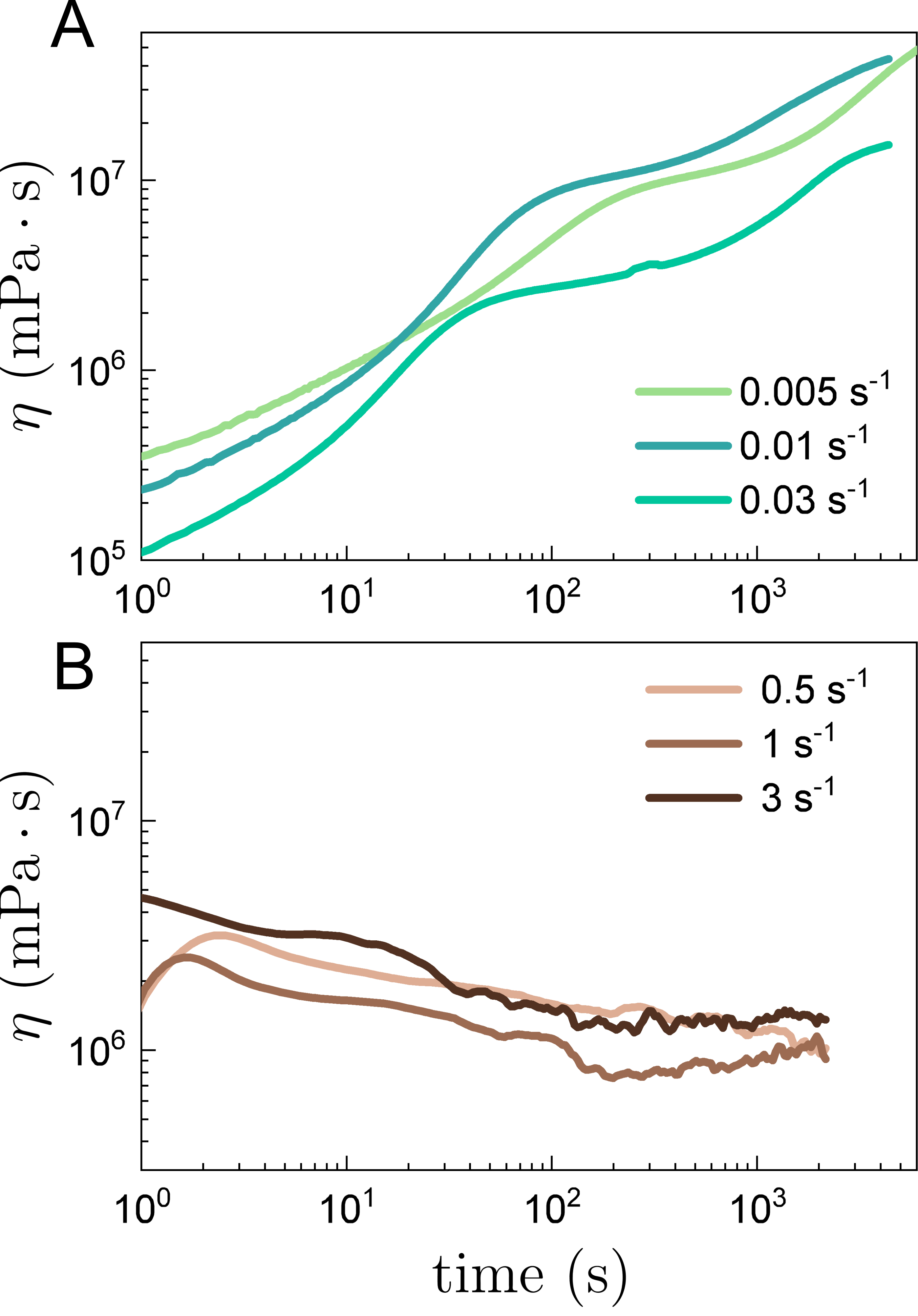}
\caption{\label{fig:timedependence} The evolution of $\eta$ of suspension with $\phi_ \mathrm{w}=40$ \% over time at fixed $\dot{\gamma}$: (A) low $\dot{\gamma}=$0.01, 0.05, and 0.1 s\textsuperscript{-1} and (B) high $\dot{\gamma}=$0.5, 1, and 3 s\textsuperscript{-1}.}
\end{figure}

\begin{figure}
\centering
	\includegraphics[width = 0.5\columnwidth]{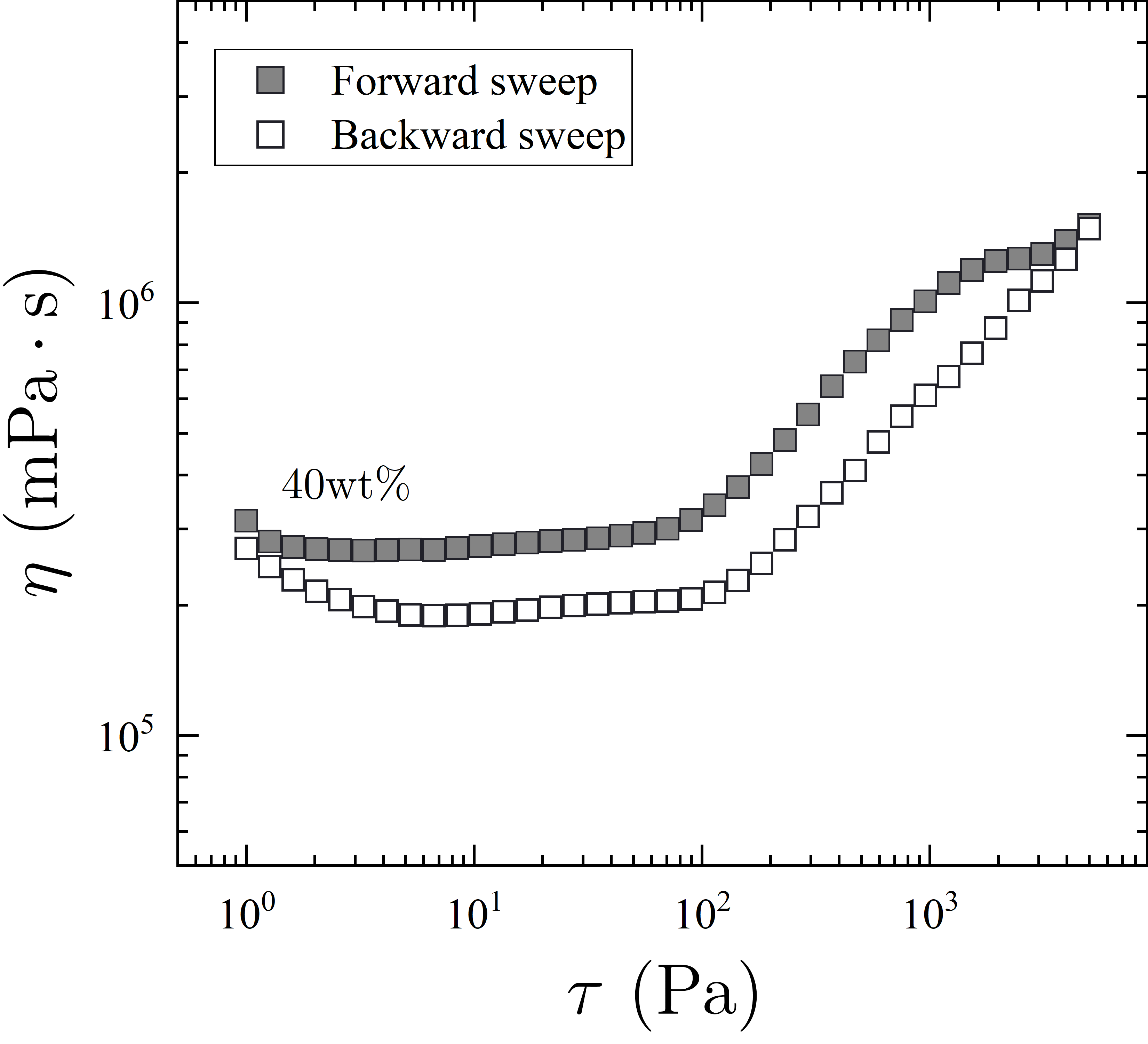}
\caption{\label{fig:OX50_OH} Rheology of unfunctionalized OX 50 fumed silica particles suspended in \textit{p}-nitrobenzalcyanoacetamide Michael-acceptor endcapped poly(propylene glycol) at weight fraction $\phi_ \mathrm{w}=40$ \%. Viscosity $\eta$ is measured by the forward (closed symbol) and backward (open symbol) sweep of applied shear stress $\tau$.}
\end{figure}

\begin{figure}
\centering
	\includegraphics[width = 0.5\columnwidth]{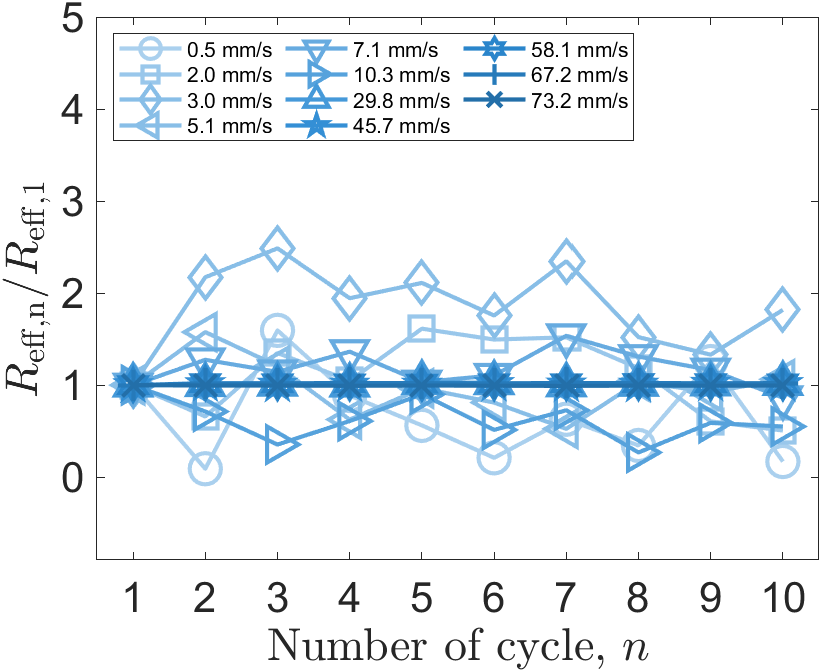}
\caption{\label{fig:OX50_PEG200_30vv} Effective impact resistance $R_\mathrm{eff,n}$ during training as a function of training cycle $n$, normalized by $R_\mathrm{eff,1}$ for the first cycle, for functionalized fumed silica OX50-SH particles and poly(ethylene glycol) (molecular weight, 200 g/mol) suspension at 30\% volume fraction.}
\end{figure}

\begin{figure}
\centering
	\includegraphics[width = 0.5\columnwidth]{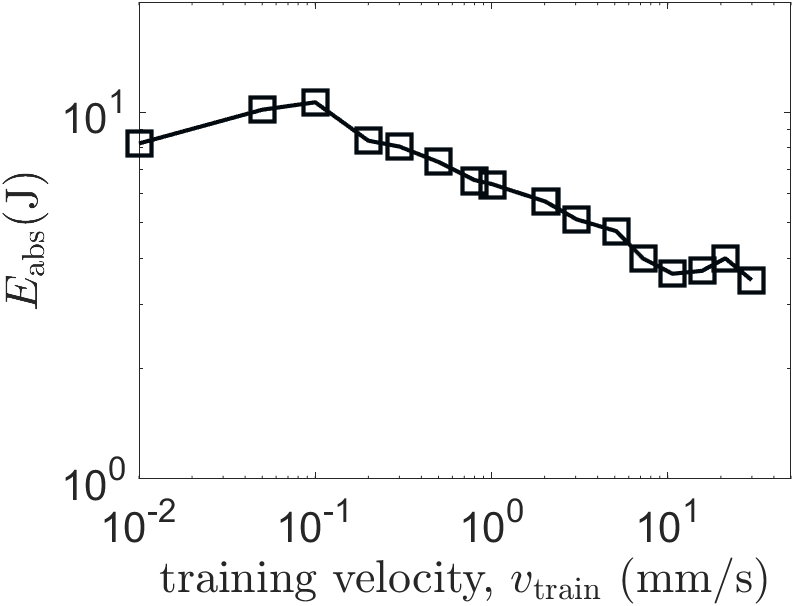}
\caption{\label{fig:E_ads_OX50SH_5mms}The amount of energy absorbed ($E_\mathrm{abs}$) during read-out with constant read-out velocity $v_\mathrm{read}=5$ mm/s after being trained at varying training velocities $v_\mathrm{train}$.}
\end{figure}

\end{document}